# Does the Planetary Dynamo Go Cycling On? Re-examining the Evidence for Cycles in Magnetic Reversal Rate

Adrian L. Melott[1], Anthony Pivarunas[2], Joseph G. Meert[2], Bruce S. Lieberman[3]

[1]University of Kansas, Department of Physics and Astronomy, Lawrence, KS, 66045;

melott@ku.edu; 785-864-3037

[2]University of Florida, Department of Geological Sciences, 241 Williamson Hall, Gainesville, FL 32611

[3]University of Kansas, Department of Ecology & Evolutionary Biology and Biodiversity Institute, 1345 Jayhawk Blvd., Lawrence, KS 66045

**Abstract**

The record of reversals of the geomagnetic field has played an integral role in the development of plate tectonic theory. Statistical analyses of the reversal record are aimed at detailing patterns and linking those patterns to core-mantle processes. The geomagnetic polarity timescale is a dynamic record and new paleomagnetic and geochronologic data provide additional detail. In this paper, we examine the periodicity revealed in the reversal record back to 375 million years ago (Ma) using Fourier analysis. Four significant peaks were found in the reversal power spectra within the 16-40-million-year range (Myr). Plotting the function constructed from the sum of the frequencies of the proximal peaks yield a transient 26 Myr periodicity, suggesting chaotic motion with a periodic attractor. The possible 16 Myr periodicity, a previously recognized result, may be correlated with 'pulsation' of mantle plumes and perhaps; more tentatively, with core-mantle dynamics originating near the large low shear velocity layers in the Pacific and Africa. Planetary magnetic fields shield against charged particles which can give rise to radiation at the surface and ionize the atmosphere, which is a loss mechanism particularly relevant to M stars. Understanding the origin and development of planetary magnetic fields can shed light on the habitable zone.

**Introduction**

The geomagnetic polarity timescale (GPTS) provides a time-series record of the Earth's magnetic field reversal frequency. Following confirmation that the reversals of the magnetic field were real and not self-reversal properties of the rocks, researchers have searched for patterns in the reversal process. Early analyses (e.g. Cox et al., 1963a) seemed to indicate a 1 Myr periodicity between polarity states, however, as more data became available (Cox et al., 1963b, 1964), the argument for a simple periodicity in the record was no longer tenable. As the word is often misused, even in the scientific literature, we will adopt the definition of periodicity as "the tendency to recur at regular intervals". Note that this only need be a tendency, not an absolutely rigid timing. Statistical tests focus on whether any observed tendency is significant, i.e. unlikely to have arisen by chance from the given background.

Later compilations, based on a more chronologically precise and detailed reversal record, led to multiple claims of a possible periodicity in the ~15-30 Myr range (Mazaud et al., 1983; Raup, 1985a). However, subsequent investigations (e.g., McFadden, 1984b; McFadden and Merrill, 1984; Lutz, 1985; McFadden et al., 1987; Lutz and Watson, 1988) questioned the methodology these investigations used and proposed that any apparent periodicity resulted from the analytical methods employed rather than geodynamic processes. Notably Raup (1985b), who initially supported claims of periodicity, agreed with Lutz (1985) that the methods and results which suggested periodicity were not robust.

Moreover, McFadden and Merrill (1984) suggested that any cyclicity should operate on the order of $10^8$ years and result from core-mantle interactions.

Still, the debate was far from settled. Stothers (1986) utilized techniques recommended by Lutz (1985) and McFadden (1984a), and suggested that the 30 My cycle was, indeed, a robust signal. Subsequent studies (Mazaud and Laj, 1991) analyzed the reversal frequency for the past 83 Ma (post-Cretaceous Long Normal) and found frequency peaks between 12.5-16.5 Myr. They argued these periodicities are real (see also Marzocchi and Mulargia, 1992). Most recently, Driscoll and Evans (2016) sought to identify superchrons in the Proterozoic paleomagnetic record, and found a ca. 200 Myr recurrence interval between superchrons. Thus, it is reasonable to conclude that the longstanding debate about periodicity in the magnetic reversal record is unresolved.

Statistical analysis of the reversal record is required to discern any periodic behavior of the magnetic field, but there are some issues that need thoughtful consideration. The common paradigm is that reversals are stochastic, independent of one another, and rare (McFadden, 1984). The fact that the reversal process is non-instantaneous and that hundreds of reversals occur in the Phanerozoic alone may belie the last two assumptions. Regardless, the reversal process has been long thought of as a renewal process, specifically Poisson, with some aberrant behaviors (McFadden, 1984a; McFadden and Merrill, 1984; Constable, 2000). McFadden (1984a) held that the sequence of reversals arising from the Poisson process is well-described by a gamma distribution with $k = 1$ and interval lengths distributed exponentially. Departing from that paradigm, Carbone et al. (2006) found that reversals can be better described as a Levy (rather than Poisson) process, and that reversal events can be correlated due to inherent 'memory' in the geodynamo. They were joined in rejecting a Poisson model and seeing correlations in reversals by Sorriso-Valvo et al. (2007). The statistical model that best describes the Earth's geodynamo remains an area of active research.

Whether or not periodicity exists in the reversal frequency is an important question, both in regard to its potential effects on the atmosphere/biosphere and for its potential in providing constraints for the origin and evolution of the geodynamo (Black, 1967; Hays, 1971; Olson, 1983; Raup, 1985a; Courtillot and Besse, 1987, McFadden, 1987; Loper et al., 1988; Valet and Vallades, 2010; Biggin et al., 2012; Wei et al., 2014; Meert et al., 2016; Gallet and Pavlov, 2016). A variety of literature has suggested that long-term cycles in extinction and/or total biodiversity exist (Raup and Sepkoski, 1982; Rohde and Muller, 2005; Lieberman and Melott, 2007, 2012; Melott and Bambach, 2011a, b, 2014; Rampino, 2015), which has triggered the desire to identify possible drivers for these long-term cycles (e.g., Medvedev and Melott, 2007; Melott et al., 2012; Rampino and Prokoph 2013; Rampino, 2015). Discovering cycles in magnetic reversals and biodiversity existing on equivalent time scales might offer a possible driver, wherein, during a reversal the magnetic field strength typically declines to very low levels which is thought to allow more cosmic rays to reach deep into the Earth's atmosphere, and even the surface at lower latitudes (Valle and Vallades, 2010; Wei et al., 2014; Meert et al., 2016). If the reversal rate was high, this especially could afford more opportunities for the penetration of cosmic rays and other disturbances in the upper atmosphere (Wei et al., 2014; Meert et al., 2016). It must be stated, however, that the increase in radiation on the ground during a reversal is not great (Glassmeier and Vogt, 2010). In particular, the rigidity due to the Earth's magnetic field varies a great deal with location, but is only of order ~ 10 GV (Herbst et al., 2013). Furthermore, it is only protons (the primary component of cosmic rays) with energy much greater than several GeV that can ionize the lower atmosphere or generate radiation on the ground (Thomas et al., 2016), although lower energy protons can ionize the stratosphere and generate ozone depletion with concomitant increases in UVB. Raup (1985a) explicitly tried to link cycles in extinction with reversal periodicity. As stated before, he soon disavowed his reversal periodicity results, but his extinction rate results remained robust under the same statistical scrutiny (Raup, 1985b; Melott and Bambach 2014, and references therein). Thus, any evidence for a potential temporal association between the periodicity of magnetic reversals and biodiversity cycles may be tenuous, at least based on current theoretical understanding. Still, atmospheric loss on the Earth during reversals has been hypothesized (Wei et al. 2014; for the case of Mars see Atri 2016). Further, planets orbiting M stars are subject to a high flux of moderate energy cosmic rays and thus loss of the protection of a magnetic field could be disastrous for any life there (Atri 2017; Atri et al. 2013). Therefore, the possibility of gaining insight into whether or not magnetic field reversals are periodic on Earth might have broader relevance.

In this paper, we re-examine the frequency of magnetic reversals using the most recent time scale and polarity data (Fig 1a, b). The majority of previous studies utilized the reversal record present in marine magnetic anomalies (MMA). MMA's are the "type section" for geomagnetic reversal polarity (Opdyke and Channell, 1996) and thus form the backbone of the GPTS. More exhaustive magnetostratigraphic work has been done on terrestrial sections, guided by quality guidelines such as the 10-point reliability index of Opdyke and Chanell (1996). In addition, calibration with astrochronology, which was limited to the Brunhes Chron (<1 Ma) during the 1980s (Opdyke and Channell, 1996), has been extended back into the Mesozoic. These improvements have not substantively altered the main pattern of reversal sequences, but even small changes in temporal resolution, identification of new polarity intervals, and an expanded database can have consequences on the statistical analysis of the sequence (McFadden et al., 1984a; McFadden and Merrill, 1984). It is important to keep in mind that the vast majority of previous statistical analyses of reversal frequency are based on ~2% of geologic time (for post-Cretaceous Long Normal analyses) to 4% of geologic time (post 160 Ma analyses). It is also important to note that a single long polarity interval (such as is found in the Cretaceous or Permo-Carboniferous) will dominate reversal frequency analyses. Long intervals are easier to detect, possibly creating a systematic bias towards them in older, less well developed reversal sequences (McFadden, 1984a).

**Material and Methods**

Magnetostratigraphic data from the latest geologic timescale (2012) was examined and a polarity and time sequence was collected from the entire Phanerozoic eon (Supplementary Table 1; Figures 1a and 1b). New results from the Devonian (Hansma et al., 2015) were utilized as no reliable data were available from the 2012 timescale. Values assigned to a polarity interval are either 0 (reverse polarity) or 1 (normal polarity). These are represented as black (1-normal) and white (0-reverse) in Figure 1a. A small number of intervals are assigned the value 0.5 when no data are available and these are shown in grey (Fig 1a). It can be seen that there are periods of rapid change (Jurassic interval) and others, as long as 40 Myr, that are of single polarity (the 'Cretaceous Long Normal' or CLN). These periods of uniform behavior of the Earth's magnetic field have other analogues – the CLN is similar to the Kiaman Reverse Superchron in the late Paleozoic or the Moyero Reverse Superchron in the Ordovician, and the more geologically recent rapid change is similar to other periods in the Devonian and Ediacaran (Hansma et al., 2015; Meert et al., 2016). These similarities seem to suggest several 'states' of the magnetic field (Meert et al., 2016; Gallet and Pavlov, 2016). The idea of states of the magnetic field has precedence, as it was previously invoked to differentiate between the Cretaceous superchron and "typical" reversal processes (Merrill and McFadden, 1994; Opdyke and Channell, 1996).

**Figure 1. Geomagnetic polarity and reversal record for the Phanerozoic.**

(a) Geomagnetic polarity back to 375.3 Ma. Intervals of normal polarity are indicated in black, reverse polarity in white, and periods of no data in grey. (On the age scale, the columns do not represent intervals of equal time.) (b) Geomagnetic reversals back to 600 Ma. The polarity is denoted by 1 for the present polarity, 0 for reversed, and 0.5 for an unknown value. (On the age scale, the columns represent intervals of equal time.) A line connects the values, so primarily black regions denote a high reversal rate, and white a region of constant value. (Note that black and white have different meanings than in [a] where they indicate polarity rather than rate of reversals.)

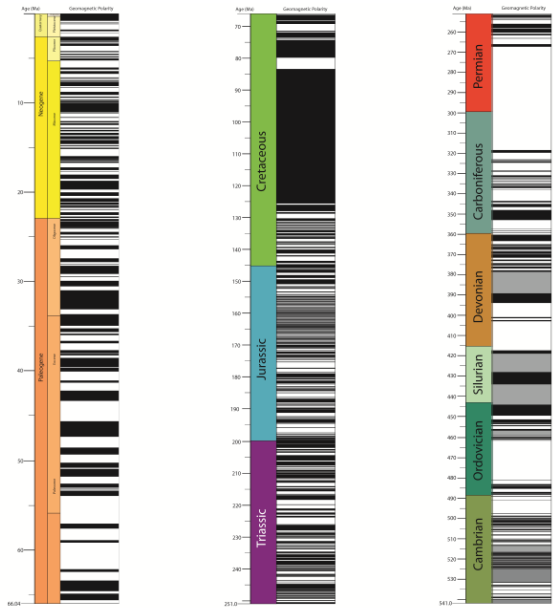

A

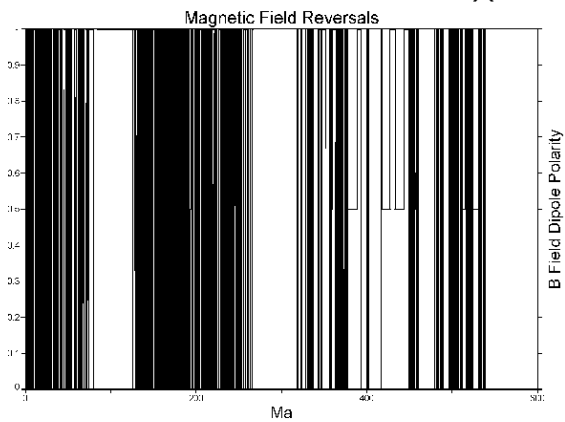

B

We note that there is no visual impression of periodicity in the signal. Analyses focused on the data back to 375 Ma (the latest Devonian), which are of higher quality than data predating the latest Devonian. The change in quality comes from the difficulty of finding unaltered sequences of rock suitable for magnetostratigraphy that are older than the latest Devonian. In particular, there has been pervasive remagnetization of Devonian and older stratigraphic sections in both North America and Europe (McCabe and Elmore, 1988; Weil and Van der Voo, 2002). As noted previously, earlier studies analyzing periodicity of magnetic reversals typically used a shorter time series (to ~160 Ma), and McFadden (1984a) pointed to data quality as one of the principal problems when analyzing these sequences.

We used Fourier analysis to identify periodicity in the reversal signal. De-trending was not applied, as there is no long-term secular trend in the data. Zero padding out to 32,768 data points was used to assure full resolution of the signal (although we do not display any results above the Nyquist frequency of the original data). No data were removed. We analyzed the data in Supplementary Table 1 using AutoSignal 1.7, a software package that has been frequently employed in time series analysis (see discussion in Melott and Bambach, 2011a). In particular, this software was used to analyze the magnetic reversal rate to rigorously test for indications of periodicity, which would be suggested by peaks in the power spectrum, and determine a significance for them. Peaks could provide evidence that a particular periodic fluctuation is stronger than would be expected by chance. However, results can be hard to interpret when several nearby frequencies, as opposed to just one, appear significant (as is discussed more fully below). Fourier analysis has been extensively used in the physical and other sciences to search for repeated patterns in time series (Cooley and Tukey, 1965; Brigham, 1988; Bloomfield, 2000; Muller and McDonald, 2002; Press et al., 2007). The method is efficacious because nearly all functions can be decomposed by Fourier analysis into a sum of sinusoids. Of course, any peaks recovered in such an analysis would probably not coincide to *precisely* timed cycles spanning much of the Phanerozoic. Also, notably, the sampling interval used herein is much smaller than the total time being considered, so this will not produce artifactual cycles, except for those with duration less than a few million years. Further discussion of these methods can be found in Muller & McDonald (2002) and Melott & Bambach (2011a, 2014).

**Results**

The analysis of power spectra of polarity revealed no interesting features, thus, as mentioned above, our analyses focused on the reversal rate; this is defined as the number of polarity changes (Fig 2) per Myr. We examined the evidence for periodicity in reversal rate using the data in Supplementary Table 1 binned into 1, 2, 4, and 8 Myr windows. 8 Myr is the largest bin size that can be used without interfering with the detection of potentially interesting cycles that might occur at ~27 Myr [Melott and Bambach, 2014; Rampino, 2015]; 1 Myr is realistically the smallest bin size amenable over the course of the Phanerozoic. Using 8 Myr bins smooths the data in a reasonable and reliable way, without producing artifacts. (Supplementary Fig 1). We analyzed the function shown in Figure 2 for periodicity, and the resulting power spectrum was plotted in Figure 3. There are three highly significant ($p < 0.001$) periodicities between 25 to 40 Myr. These peaks specifically correspond to periods of length 25.5, 29.4, and 39.4 Myr. However, the meaning of these peaks taken individually may be suspect due to their close proximity. That is to say, waves of nearby frequencies can interfere constructively and destructively to produce a pattern reflecting the sum and difference of the individual amplitudes. The actual function based on the sum of these three significant peak frequencies in Figure 3 is plotted in Figure 4. The original data are shown as points, and the reconstruction as a solid line. The three peaks interfere and produce a wave packet, strongest from 120 to 220 Ma. The validity of the choice of peaks is supported by the resemblance of peaks in the reconstructed function to the original one in Figure 4. It can be seen that they combine to produce a pattern that waxes and wanes as the phase difference of the waves changes with time. This variability (waxing and waning) explains the different results found in past research based on different time sections – particularly the elusive 30 Myr cycle mentioned previously.

**Figure 2. Geomagnetic reversal rate back to 375 Ma.**

Analyses binned using a sliding window of 8 Myr (see text). This binning not only defines the rate, but also smooths the data. Other sliding windows produced the same general structure, with less smoothness.

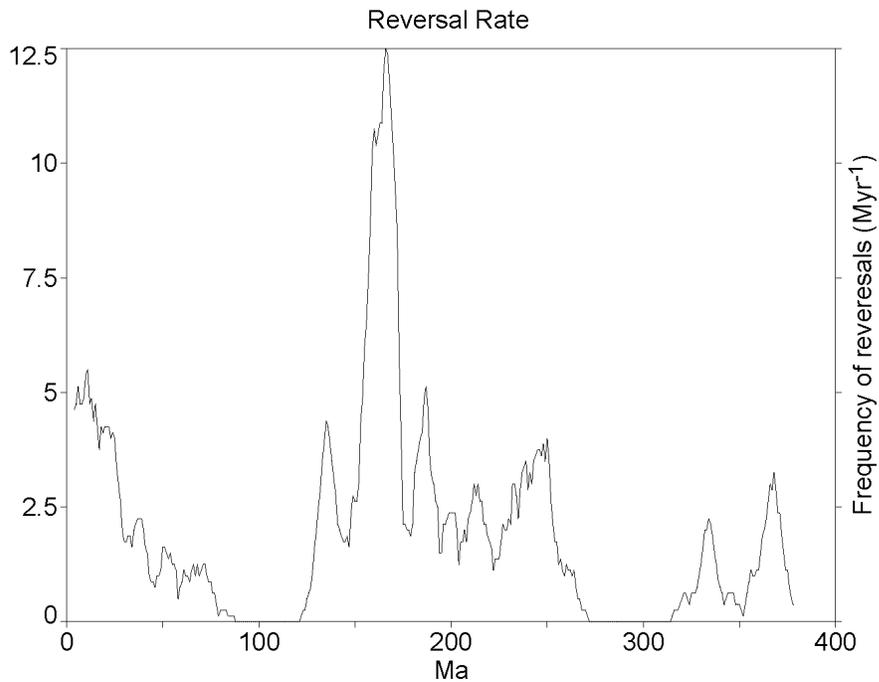

**Figure 3. Fourier frequency spectrum.**

Shown for the reversal rate function defined in Figure 2. The power of the spectrum is plotted on the y-axis, while frequency is plotted on the x-axis. Highly significant peaks appear in the period range of 25-40 Myr (which correspond to a frequency range of .04 $Myr^{-1}$ to .025 $Myr^{-1}$) and ~ 16 Myr (which correspond to a frequency range of about .06 $Myr^{-1}$). Nearly parallel lines correspond to confidence levels of .95, .99, and .999.

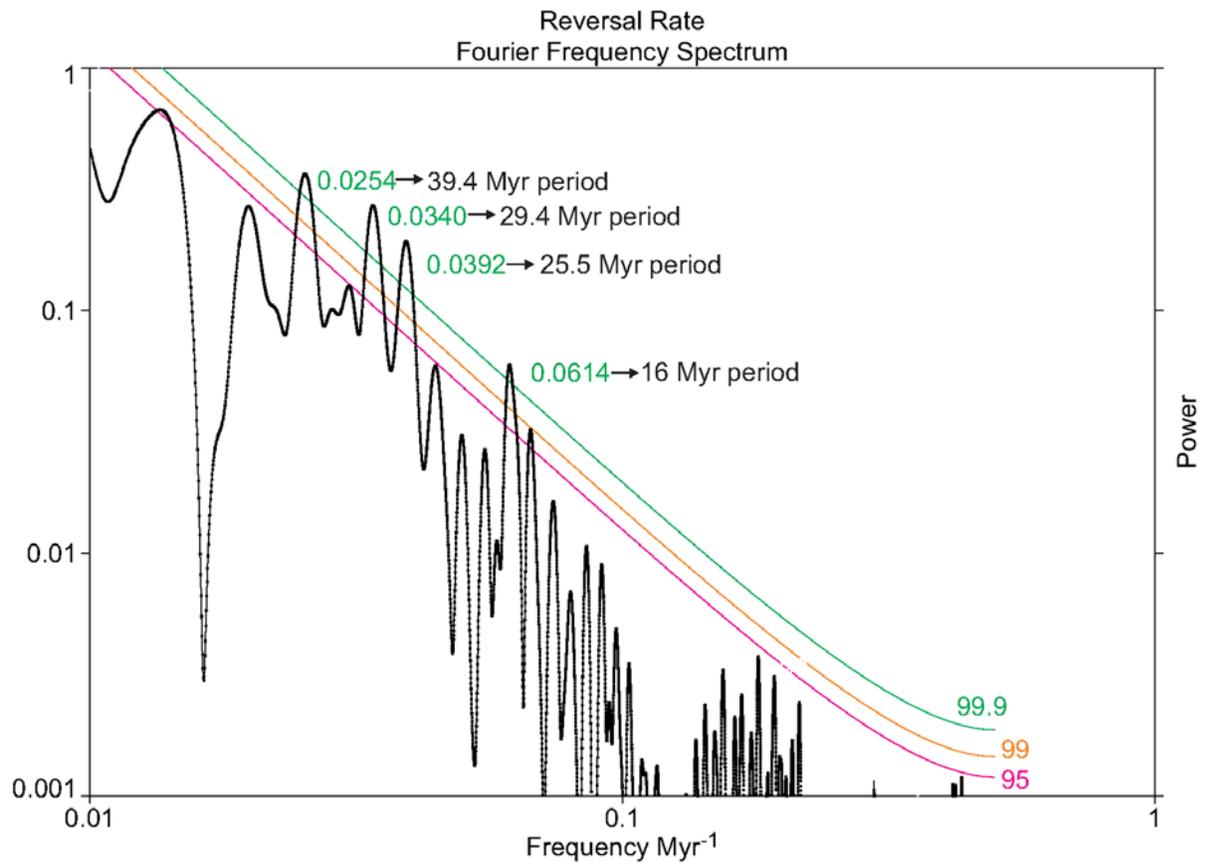

**Figure 4. The function constructed from the three nearby and significant peaks in the Fourier frequency spectrum.**

Plotted on a reversal frequency versus time graph and shown as a solid line. The wave packet that results is strongest from 150 Ma – 250 Ma. The dotted line represents the 8 Myr sliding window reversal rate.

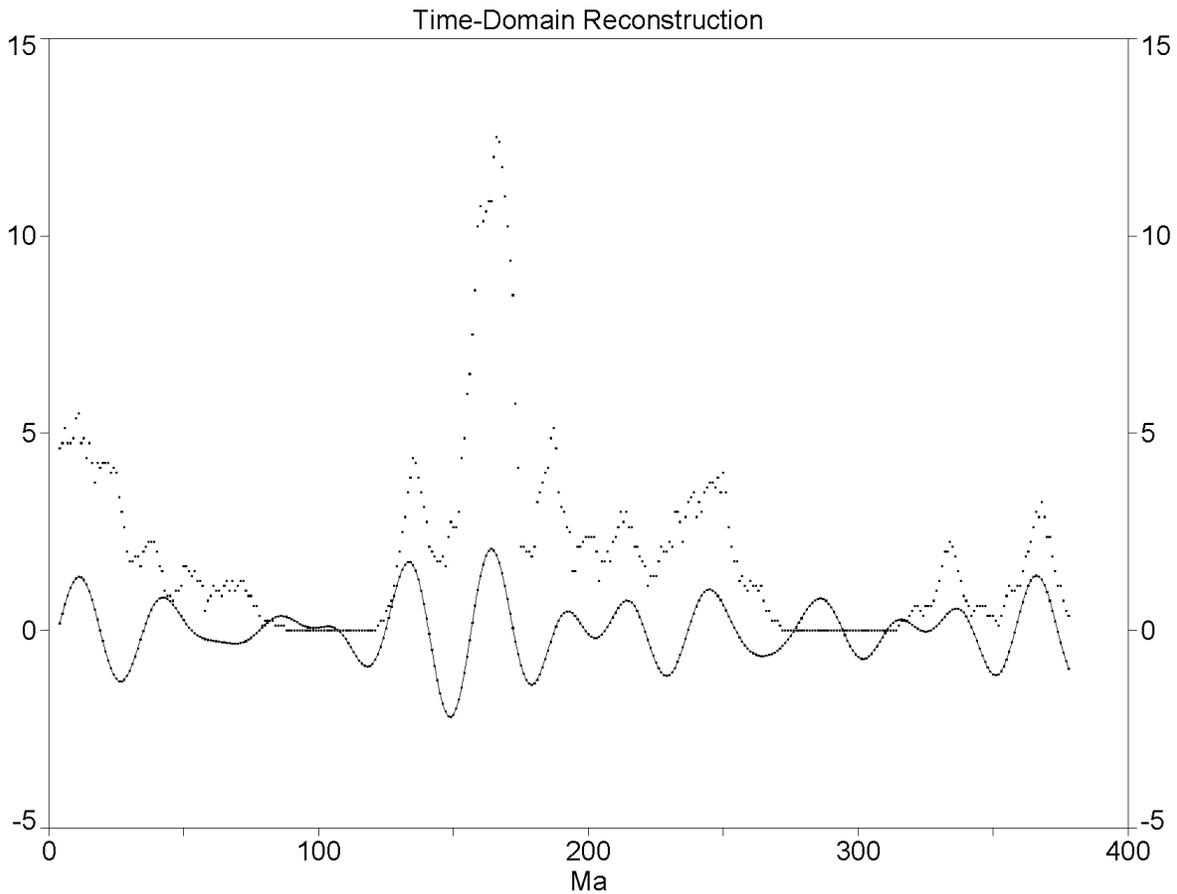

In addition, another significant peak in Figure 3 corresponds to a period of 16.3 Myr. This peak is less proximate to the other three, and therefore may be more meaningful individually. It is discussed more fully below. Notably, the multiplicity of peaks recovered by Fourier analysis suggests that there is no single cycle in the dynamo of the Earth's magnetic field, although there could be several different cycles operating on different time scales that interact in a complex way.

**Discussion**

Our results generally confirm those of Lutz (1985), who found that evidence for periodicity was affected by the waxing and waning of the magnetic signal over the last 160 Myr. This long-term variation is caused by the interaction of three peaks (Fig 4). The geodynamo may have been active for over 4 billion years (Tarduno et al, 2015), and thus unfortunately our result is only based on ~9 percent of the possible total record so clearly more information is needed. However, as mentioned above, other studies utilized even less of the total record (primarily due to the quality of the record available for analysis at the time). Thus, full appreciation of periodicity in the magnetic reversal record may well depend on the difficult task of reliably extending that record deep in the pre-Cambrian. Still, some general conclusions are possible.

In particular, our results provide no evidence at this time for simple periodicity. Instead, we find multiple significant components that interact in a complex way. Three power spectral peaks that we describe interfere to produce an episodic 26 Myr signal as seen in Figure 4. Together, they produce an interesting pattern, variable over the interval examined. It is possible that with these we are seeing a transient periodicity reflective of a semi-periodic saddle in a chaotic process (e.g. Schaffer et al. 1993). Since nearly any function can be decomposed into a sum of periodic components, usually, the question is formulated as whether any one component dominates, in which case the series is usually described as periodic. At least at this time, the window portrayed in Figure 4 is suggestive of, but does not demonstrate, a transient 26 Myr periodicity which could be repeating on a ~200 Myr timescale. Such behavior is not uncommon in chaotic systems (Kendall et al., 1993). We do note that this period is close to that noted for mass extinctions and extinction rate in general (Raup and Sepkoski 1984; Melott and Bambach 2013), although we do not yet see a likely physical connection.

It is conceivable that phenomena operating at three different time periods and possibly pertaining to different parts of the inner earth interfere, but there is no statistical evidence to indicate that is the case. The inner core, outer core, and core-mantle boundary (CMB) are understood to be significant in producing the geodynamo – yet reliable specifics as to their behavior/interactions are enigmatic. Perhaps these interactions result in different "states" of the magnetic field--hyperactive (>4-6 R/Myr) or—stable (2-3 R/Myr) each containing their own periodicity, which is obscured when the entire sequence is analyzed. The wave packet appearance at ca. 150 Ma – 250 Ma encompasses several longer period cycles proposed (Merrill and McFadden, 1984; Driscoll and Evans, 2016).

Our ~16 Myr periodic signal has been seen before in multiple analyses of the reversal record (see introduction). For instance, Mazaud and Laj (1991) and Marzocchi and Mulargia (1992) also described a ~ 15 Myr periodicity (although they only examined the past 85 Ma). As mentioned earlier, deep mantle/core processes are genetically linked to the geodynamo. Finding core-mantle boundary processes which occur at our (and others) ~ 16 Ma timescale helps show the robust nature of this result. Notably, periodicity of deep mantle plumes such as those beneath Iceland and Hawaii on the order of ~15 Myr was suggested by Mjelde and Failede (2009), Madrigal et al. (2016) and Mjelde (2016). As these plumes are hypothesized to arise from the D" layer just above the core-mantle boundary, some linkage of the magnetic reversal record with mantle plumes should be expected (Larson, 1991a). The natural question here is: does one drive the other, or are they simply part of the same process? Put more exhaustively: do energy perturbations in the lowest mantle associated with the genesis of a deep mantle plume drive reversal "states", or is heat flux at the core-mantle boundary (CMB) from the (possibly periodic) reversal process manifested as periodic plume behavior (Madrigal et al., 2016)? These would correspond to a top down or bottom up mechanism, respectively. A bottom up mechanism, favored by both Larson and Olson (1991a) and Mjelde and Faleide (2009), where heat from the core periodically transitions to the mantle, thereby causing mantle plumes, should have an effect on the reversal process – which explains the similar periodicity in the magnetic reversal record. The 10-20 Myr periodicities documented by Mjelde and Failede (2009), Mjelde (2016) and Madrigal et al. (2016) extend the periodic record back to the Jurassic Period. Torsvik et al. (2014) argue that the large low shear velocity provinces (hereafter LLSVPs) beneath Africa (Tuzo LLSVP) and the Pacific Ocean (Jason LLSVP) are long-lived features of the planet. If the heat release is being pulsed on a 10-20 Myr time scale and dynamically related to the Earth's dynamo, then it is not surprising that we see that periodicity when the reversal record is examined back into the Paleozoic.

We note that the CMB is an extensive, heterogeneous thermal boundary layer. Plume activity, even at two disparate points on the Earth's surface, may not be representative of general behavior of the entire core-mantle boundary. A more detailed explanation linking reversals and heat flux at the CMB can be found in Larson (1991a). Larson (1991b) also proposed that this mechanism caused the CLN – wherein a large pulse of heat loss from the core led to both no reversals and a superplume. The lack of reversals during this time strongly suggests a relationship between heat loss and reversal frequency. The scale of effect of a superplume is clearly much greater than that of the "pulsing" noted by Mjelde and Failede (2009). However, since even normal plume activity may be responsible for the removal of much if not all of the heat lost from the core into the lower mantle (Lay et al., 2008), the coincidence of timescales for both plumes and reversals cannot be rejected as meaningless based on calling the proposed CLN connections a unique occurrence. If unusual thermal dynamics at the CMB resulting in a superplume can lead to a superchron in the reversal record, then normal thermal dynamics at the CMB could lead to a recognizable signal in the reversal record. Olson and Amit (2015) also agree with this conclusion, calling a link between the two "logical".   Bearing these caveats in mind, we believe the ca. 16 Myr period for reversals considered together with ca. 15 Myr "pulsing" of plumes is a new piece of evidence for field reversal-plume connections.  Ultimately, the spatial and temporal association of these two geodynamic phenomena is inescapable, yet it could still of course be coincidental.   It does bear mentioning that plate tectonics would not be expected to operate in the same manner, or even exist, on other planets that might possess life, so it is impossible to generalize the relevance of this Earth-based mechanism to exoplanets.

## Conclusions

The rate of geomagnetic reversals could potentially be relevant to life on Earth and its extinctions (Wei et al. 2014), especially since a magnetic reversal occurring during transit of a dense interstellar cloud could be catastrophic (Pavlov et al. 2005). The question of reversal probability will be even more relevant for the long term potential for life to survive (or the type of life that survives) around M stars (Atri et al. 2013; Atri 2017) due to intense flare activity. Fourier analysis of the magnetic reversal record reveals multiple significant periodic components, some of which seem to interact in a complex way and thus may be spurious, but one periodic component may actually be significant and possibly finds physical explanation. In particular, the three neighboring peaks in the magnetic reversal rate power spectrum that conform to peaks at periods of length 25.5, 29.4, and 39.4 Myr and produce an inconsistent 26 Myr signal probably do not reflect true periodicity in reversal rate. Instead, transient periodicity is a better fit to these aspects of the data. By contrast, we posit that the peak in reversal rate at a period of ~ 16 Myr could be real and may be correlated with plume-related activity via heat-exchange processes at the CMB near the LLSVP regions. Ultimately, more magnetic reversal data from before 375 Ma will help address the resiliency of these results.

## Acknowledgments

ALM is grateful for grant support from NASA Exobiology grant NNX14AK22G. BSL was supported by NSF DBI-1602067. We thank Bruce Buffett for comments on a previous version of this manuscript.

**Figure Captions**

**Figure 1. Geomagnetic polarity and reversal record for the Phanerozoic.**

Geomagnetic polarity back to 375.3 Ma. Intervals of normal polarity are indicated in black, reverse polarity in white, and periods of no data in grey. (On the age scale, the columns do not represent intervals of equal time.) (b) Geomagnetic reversals back to 600 Ma. The polarity is denoted by 1 for the present polarity, 0 for reversed, and 0.5 for an unknown value. (On the age scale, the columns represent intervals of equal time.) A line connects the values, so primarily black regions denote a high reversal rate, and white a region of constant value. (Note that black and white have different meanings than in [a] where they indicate polarity rather than rate of reversals.)

**Figure 2. Geomagnetic reversal rate back to 375 Ma.**

Analyses binned using a sliding window of 8 Myr (see text). This binning not only defines the rate, but also smooths the data. Other sliding windows produced the same general structure, with less smoothness.

**Figure 3. Fourier frequency spectrum.**

Shown for the reversal rate function defined in Figure 2. The power of the spectrum is plotted on the y-axis, while frequency is plotted on the x-axis. Highly significant peaks appear in the period range of 25-40 Myr (which correspond to a frequency range of .04

Myr$^{-1}$ to .025 Myr$^{-1}$) and ~ 16 Myr (which correspond to a frequency range of about .06 Myr$^{-1}$). Nearly parallel lines correspond to confidence levels of .95, .99, and .999.

**Figure 4. The function constructed from the three nearby and significant peaks in the Fourier frequency spectrum.**

Plotted on a reversal frequency versus time graph and shown as a solid line. The wave packet that results is strongest from 150 Ma – 250 Ma. The dotted line represents the 8 Myr sliding window reversal rate.

**Supplementary Figure 1**

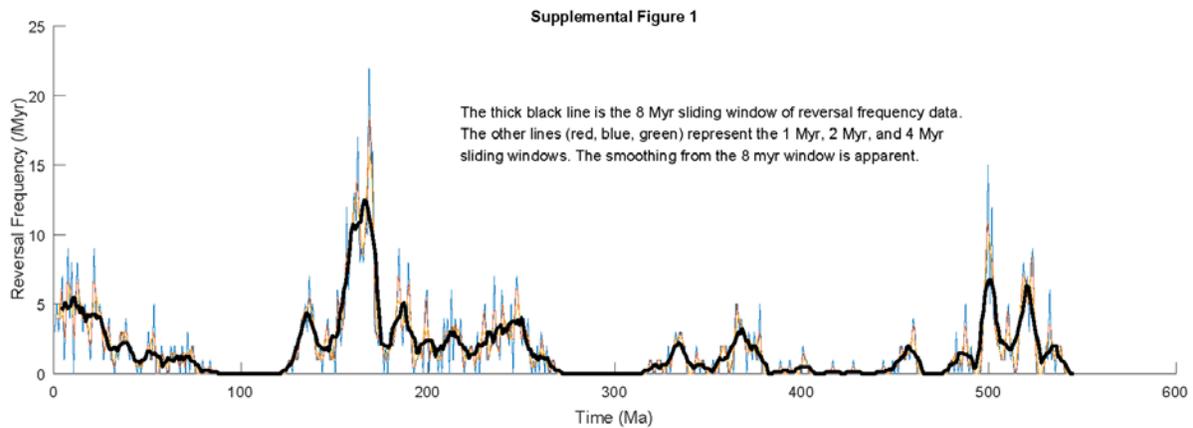